\documentclass[12pt,a4paper]{article}
\usepackage{graphicx}
\usepackage{rotating}
\usepackage{lscape}
\usepackage{comment}
\usepackage[square,sort,comma,numbers]{natbib}

\usepackage{url}
\usepackage{xcolor}
\usepackage{amsmath}
\usepackage{prodint}

\usepackage{authblk}

\title{Integrating relative survival in multi-state models -- a non-parametric approach}
\author[1]{Damjan Manevski}
\author[2]{Hein Putter}
\author[1]{Maja Pohar Perme}
\author[2]{Edouard F.~Bonneville}
\author[3]{Johannes Schetelig}
\author[2]{Liesbeth C.~de Wreede}
\affil[1]{Institute for Biostatistics and Medical Informatics, Faculty of Medicine, University of Ljubljana, Vrazov trg 2, 1000 Ljubljana, Slovenia}
\affil[2]{Leiden University Medical Center, Einthovenweg 20 2333 ZC Leiden, the Netherlands}
\affil[3]{Medical Clinic I, University Hospital, Fetscherstrasse 74, 01307 Dresden, Germany}

\usepackage{listings}
\begin{document}
\maketitle


\begin{abstract} 

Multi-state models provide an extension of the usual survival/event-history analysis setting. In the medical domain, multi-state models give the possibility of further investigating intermediate events such as relapse and remission. In this work, a further extension is proposed using relative survival, where mortality due to population causes (i.e.~non-disease-related mortality) is  evaluated. The objective is to split all mortality in disease and non-disease-related mortality, with and without intermediate events, in datasets where cause of death is not recorded or is uncertain. To this end, population mortality tables are integrated into the estimation process, while using the basic relative survival idea that the overall mortality hazard can be written as a sum of a population and an excess part. Hence, we propose an upgraded non-parametric approach to estimation, where population mortality is taken into account. Precise definitions and suitable estimators are given for both the transition hazards and probabilities. Variance estimating techniques and confidence intervals are introduced and the behaviour of the new method is investigated through simulations. The newly developed methodology is illustrated by the analysis of a cohort of patients followed after an allogeneic hematopoietic stem cell transplantation. The work is also implemented in the \texttt{R} package \texttt{mstate}.


\end{abstract}

\section{Introduction}

Multi-state models provide a framework for simultaneously analyzing competing events and sequences of events. In the medical field, these models help to study the impact of intermediate events on the prognosis of patients, and allow to estimate separate probabilities of death with and without the intermediate event at multiple time horizons. Interest can also be in distinguishing between deaths due to the studied disease and its treatment (excess mortality) and deaths due to other (population) causes. This is of particular interest in an older patient population where the risk of dying due to other causes is high and may considerably contribute to the overall proportion of deaths. However, cause of death is often not reported \cite{mariotto2014}, unreliable \cite{begg2002, percy1981} or cannot unequivocally be attributed to the disease or other causes.

Our motivating example comes from the study of outcomes for patients with myelodysplastic syndromes (MDS) or secondary acute myeloid leukemia (sAML) after an allogeneic hematopoietic stem cell transplantation (alloHCT) \cite{leukemia2019}. Patients with these indications represent the oldest major patient population referred to alloHCT with a median age of 58 years at alloHCT (registry data from 2012). These patients are at significant risk for two competing failures: relapse of the underlying disease and non-relapse mortality (NRM), which for a large part is due to the transplantation and pre-treatment. The occurence of relapse leads to a very poor prognosis. We investigated the contribution of population mortality to both death after relapse and NRM. This enabled us to estimate the probability of excess NRM, which especially for older patients and for long-term outcomes may provide a better estimate for treatment-related mortality than all NRM. The current paper discusses and extends the model introduced in that paper.

When reliable information on causes of death is not available in the medical data, one needs to address this issue using external data. This is the core idea of the relative survival field \cite{mppnet}: the observed data is  merged with general population mortality tables \cite{HMD} to indirectly enable the estimation of cause-specific hazards and probabilities \cite{ederer, hakmethod}. In this work, we extend the ideas of the relative survival methodology to a Markov time-inhomogeneous multi-state model where we focus on non-parametric estimation of transition hazards and probabilities. We provide a theoretical foundation, a study of behaviour of the methods under different scenarios and a software implementation in \texttt{R} that makes the methodology readily available to other users.

These measures have also been the focus of the relative survival analysis field, which has seen considerable development in the last decade \cite{mppnet, Sasieni2016, pavlic2018}. With the generalisation to multi-state models, a new level of complexity is added. The main contribution of this paper is to discuss which measures may be of main interest in this setting and how to estimate them. We propose estimators for cumulative hazards and transition probabilities and study their properties, including their (asymptotic) variances. To the best of our knowledge, this is the first extension of relative survival to non-parametric estimation in a Markov multi-state model, whereas other semi-parametric \cite{huszti2012, belot2011, gillaizeau2017} or parametric \cite{crowther2017, weibull2021} approaches have already been proposed. An extensive simulation study of properties of the estimators in terms of bias, standard error and coverage probability for confidence intervals has thus been performed. The simulation study is also the first (as to our knowledge) to extensively investigate the behaviour of different confidence interval methods for non-parametric multi-state models.


The article is structured as follows: in Section \ref{theory} we present the main theoretical grounds of our work. In Section \ref{sim} we present simulation results by which this new approach is evaluated. In Section \ref{example} we will reanalyse the motivating dataset to illustrate the behaviour and interpretation of the new model. This also serves as a step-by-step overview of how such an analysis can be performed in \texttt{R} using the newly developed \texttt{R} code. In Section \ref{discussion} we discuss the main findings of the article and provide conclusions.	


\section{Extended multi-state model integrating relative survival}\label{theory}

We start by introducing the theoretical background of multi-state models and relative survival needed for our work in Sections \ref{mstatesec} and \ref{relsurvsec}. We then present our proposed extension in Section \ref{mstaterelsurvsec}.

\subsection{General multi-state model} \label{mstatesec}

A multi-state model is a stochastic process $(X(t), t \in [0, \tau))$ with a finite state space $S$ of size $K$, $\tau \in {\rm I\!R}^+$, where the sample paths are assumed to be right-continuous, i.e.~$X(t+) = X(t)$. Thus, $X(t)$ represents the state at which the process (or a given individual) is present at time $t$. A transition from state $h$ to state $j$ takes place when an individual at risk in state $h$ experiences an event that makes them enter state $j$.  


By assuming $X(t)$ is a Markov process we can define transition probabilities in the following way

\begin{equation}
P_{hj} (s, t) = \text{Prob} ( X(t) = j | X(s) = h), 
\label{phj}
\end{equation}
for every possible combination of states $h, j \in S$; $s, t \in [0, \tau), s\leq t$, where the probability does not depend on the history up to (but not including) time $s$. 


From this we can define transition hazards (or transition intensities) as the derivative of $P_{hj} (s, t)$ with respect to $t$ (which we assume to exist), evaluated at $s=t$

$$\lambda_{hj} (t) =  \lim_{\Delta t \rightarrow 0} \frac{P_{hj} (t, t + \Delta t)}{\Delta t};$$ 
the cumulative hazard is then defined as
$$\Lambda_{hj} (t) = \int\limits_0^t \! \lambda_{hj} (u) \, \mathrm{d}u   .$$


Apart from the definition in formula \eqref{phj}, transition probabilities can also be defined in a matrix form

\begin{equation}
\mathbf{P}(s, t) = \Prodi_{(s, t]} (\mathbf{I} + \boldsymbol{\lambda} (u) \mathrm{d}u) .
\label{pt_matrix}
\end{equation}
where $\mathbf{P}(s, t)$ represents a square matrix of size $K \times K$. In such a matrix, every entry represents the transition probability of being in a given state at time $t$ conditional on being in a given state at time $s$. Similarly, $\boldsymbol{\lambda} (t)$ is a matrix of size $K \times K$ containing the hazards $\lambda_{hj} (t)$ for all possible transitions $h \rightarrow j, h \neq j$; whereas diagonal values are defined as $\lambda_{hh} (t) = - \sum\limits_{j \neq h} \lambda_{hj} (t)$.


%

\begin{figure}[ht!]
\begin{center}
\includegraphics{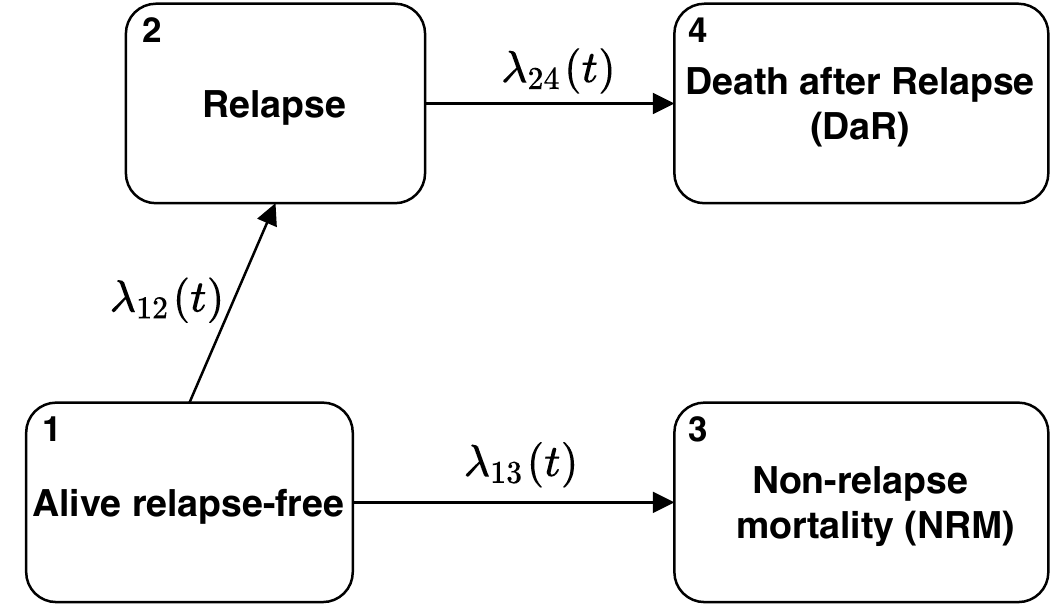}
\caption{Transition diagram for the basic multi-state model.}
\label{transmat}
\end{center}
\end{figure}

Both transition hazards and probabilities are of key interest for our work. In Figure \ref{transmat} we introduce the multi-state model which will be used as the leading example in the article. In this model, all patients with a certain disease are alive at time zero and have no change of their underlying disease status (e.g.~relapse). From the starting state they can then experience two competing risks: relapse (an intermediate event) or death (i.e.~non-relapse mortality, NRM; an absorbing state).  Death can also occur after relapse. We call this state death after relapse (DaR) which is also an absorbing state. Thus, the model in Figure \ref{transmat} is the usual illness-death model where the death state is split into two different states in order to calculate separate probabilities of NRM and DaR (which sum up to total mortality). Although the intermediate event considered is relapse, any other relevant event could be considered, e.g.~recurrence, progression, adverse event, additional treatment, or recovery. In the following, we assume that the censoring times for relapse and death are equal, that all relapses before the individual censoring times are recorded in the dataset and that the event times are observed exactly.

This model contains the key features of a multi-state model: combinations of competing risks and series of events. All theory and the implementation in software discussed below are also valid when the model is extended with additional states. They also hold when individuals start in different states or are only observed after a certain timepoint (left-truncated observations).

We now turn to estimation. The Nelson-Aalen estimator is the common non-parametric estimator used for the cumulative transition hazard:
\begin{equation}
\widehat{\Lambda}_{hj} (t) = \int\limits_0^t \! \frac{\mathrm{d} N_{hj} (u)}{ Y_{h} (u)},
\label{NA}
\end{equation}
where $N_{hj} (u)$ represents the counting process that gives the number of $h \rightarrow j$ transitions in the time interval $[0, u]$, whereas $Y_{h} (u) = \sum\limits_{i=1}^n Y_{h, i} (u)$ gives the size of the at-risk set, i.e.~the number of individuals present in state $h$ just before time $u$ ($Y_{h, i} (u)$ indicates whether the i$^{th}$ individual is present in state $h$ just before time $u$).

For estimating the variance of the Nelson-Aalen estimator the Greenwood estimator is used, as commonly suggested in the literature \cite{mstate2010, Beyersmann2010, Beyersmann2017, Beyersmann2019}:

\begin{equation}
\widehat{\text{Var}}_{G} (\widehat{\Lambda}_{hj} (t) ) = \int\limits_0^t \! \frac{ \mathrm{d} N_{hj} (u)}{Y_{h} (u) (Y_{h} (u) - \mathrm{d} N_{hj} (u))}. 
\label{var_greenwood}
\end{equation}

For estimating transition probabilities the Aalen-Johansen estimator is standardly used; it is obtained from formula \eqref{pt_matrix} by replacing $ \boldsymbol{\lambda} (t)$ with the corresponding estimator $\boldsymbol{\hat{\lambda}} (t)$:

\begin{equation}
\mathbf{\hat{P}}(s, t) = \Prodi_{(s, t]} (\mathbf{I} +  \boldsymbol{\hat{\lambda}} (u) \mathrm{d}u),
\label{transprob}
\end{equation}
or written in discrete form: $\mathbf{\hat{P}}(s, t) = \displaystyle \prod_{u \in (s, t]} (\mathbf{I} +  \Delta \boldsymbol{\widehat{\Lambda}} (u))$, where $u$ indicates all event times in $(s, t]$ in the dataset.

We point out that both the Nelson-Aalen and Aalen-Johansen estimator are consistent when the multi-state model is Markov. For a non-Markov process the Nelson-Aalen estimator is also consistent \cite{datta}. We also note that by using the landmark Aalen-Johansen estimator, each $P_{hj} (s, t)$ can be estimated consistently \cite{putter2018}.

For estimating the variance of transition probabilities we can again use the Greenwood estimator. We denote it as $\widehat{\text{Var}}_G ( \mathbf{\hat{P}}(s, t) )$, for which an exact definition is available in \cite[p. 294]{andknjiga} and \cite{mstate2010}. 


Here we have provided only the basic tools from multi-state models that are needed for this work; a more thorough and theoretical introduction can be found in \cite{andkeid02, putter, handbook2014}.

\subsection{Relative survival} \label{relsurvsec}

We will now introduce the main ideas of relative survival which are relevant for
this work, using the basic survival setting where only one event (death) is reported in the data. Suppose we are interested in distinguishing between two causes of death -- death due to the disease and its treatment and due to other (population) causes. To make this distinction we use population mortality tables as an additional source of information. Such mortality tables give hazards for every individual based on demographic covariates which have to be present both in the dataset and in the mortality tables (e.g.~age, sex, year, country) to allow for proper matching. This enables us to evaluate the amount of deaths due to other (population) causes in the diseased population.

\begin{figure}[ht!]
\begin{center}
\includegraphics[height=16mm]{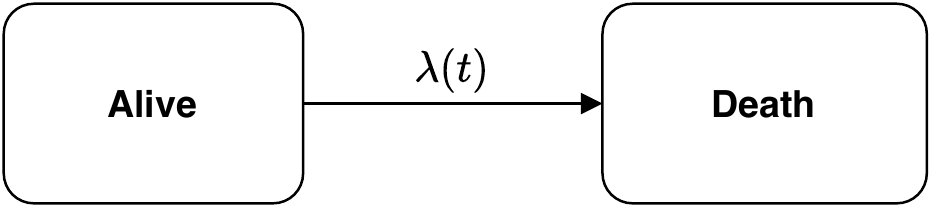} \hspace{3mm}
\includegraphics[height=16mm]{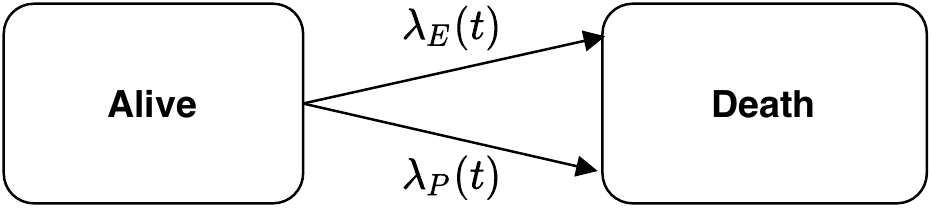}
\end{center}
\caption{Illustration of the additive model given in formula \eqref{hazardAditivniModel}.}
\label{relsurvfig}
\end{figure}

This information has to be appropriately incorporated in a model. We assume that the (observed) hazard of dying $\lambda (t)$ is the sum of two (unobserved) hazards, the hazard of dying because of the disease (i.e.~the excess hazard, denoted as $\lambda_{E}(t)$) and the hazard of dying because of other causes (the population hazard, denoted as $\lambda_{P}(t)$). This assumption characterizes the additive model:
\begin{equation}
\label{hazardAditivniModel}
\lambda(t) = \lambda_{E}(t) + \lambda_{P}(t).
\end{equation}
The model is illustrated in Figure \ref{relsurvfig}. In the interest of clarity we give an exact definition of these hazards. Let $T_E$ be the time to excess death and $T_P$ time to population death (i.e.~death from other causes). We denote the minimum of these two times as $T$, the time of death. Time $T$ can be also subject to a censoring time $T_C$ which we assume to be non-informative. Hence, $\widetilde{T} = \text{min} (T, T_C)$ is observed, together with the event indicator at $\widetilde{T}$ (event or censoring). Then the hazard definition follows
$$\lambda (t) = \lim_{\Delta t \to 0} \frac{P(t < T \leq t + \Delta t | T > t)}{\Delta t};$$
$$\lambda_{E}(t) = \lim_{\Delta t \to 0} \frac{P(t < T_E \leq t + \Delta t | T > t)}{\Delta t}; \hspace{15pt} \lambda_{P}(t) = \lim_{\Delta t \to 0} \frac{P(t < T_P \leq t + \Delta t | T > t)}{\Delta t}.$$

The two hazards $\lambda_{E}(t)$ and $\lambda_{P}(t)$ are referred to as cause-specific hazards \cite{tsiatiscomp} in analogy to the standard competing risks model where both causes of death are observed. 

It is important to note that in the definition of $\lambda_{E}(t)$ and $\lambda_{P}(t)$ we condition on $\{T > t\}$. This means that we condition on the observed time and as put in \cite{andersenKeiding} we “stick to the real world". Another possibility would be to condition on $\{T_E > t\}$ and $\{T_P > t\}$, respectively, which is commonly done when defining net survival \cite{mppnet}. A more in-depth study of hazard definitions is provided in \cite[Section~7]{pavlic2018}, \cite{mppnet, Sasieni2016}. We further assume that the population of interest is small compared to the general population, i.e., that removing the subpopulation of interest from population life tables would have negligible effect on $\lambda_P$.

$\Lambda (t) = \int_0^t \! \lambda(u) \mathrm{d}u$ can be estimated using the Nelson-Aalen estimator.  $\lambda_P (t)$ is estimated using mortality tables: for the i$^{th}$ individual we can obtain $\lambda_{Pi} (t)$ based on demographic covariate values. An estimator for the cumulative version of $\lambda_P (t)$ is then defined as
\begin{equation}
\widehat{\Lambda}_{P}(t)  = \int\limits_0^t \! \frac{\sum\limits_{i=1}^n Y_i (u) \mathrm{d}\Lambda_{Pi} (u) }{Y (u)},
\label{lambdaP}
\end{equation}
based on $n$ individuals in the dataset. The definition of $\widehat{\Lambda}_{P}(t)$ corresponds to the definition of $\Lambda_{P}(t)$ when written with respect to the covariate distribution, see formula (1) in the Supplementary material where we have also provided a derivation.

The estimator for the cumulative excess hazard $\Lambda_E (t)$ is defined as the difference of the estimators for the observed and population hazard, that is,
\begin{equation}
\widehat{\Lambda}_{E} (t) = \int\limits_0^t \! \frac{\mathrm{d} N  (u)}{ Y (u)} - \int\limits_0^t \! \frac{\sum\limits_{i=1}^n Y_i (u) \mathrm{d} \Lambda_{Pi} (u) }{Y (u)},
\label{lambdaE}
\end{equation}
where $N(u)$ represents the number of events in $[0, u]$.


\subsection{Extended multi-state model integrating relative survival}\label{mstaterelsurvsec}

\subsubsection{Estimation of hazards and probabilities}
	
\begin{figure}[ht!]
\begin{center}
\includegraphics{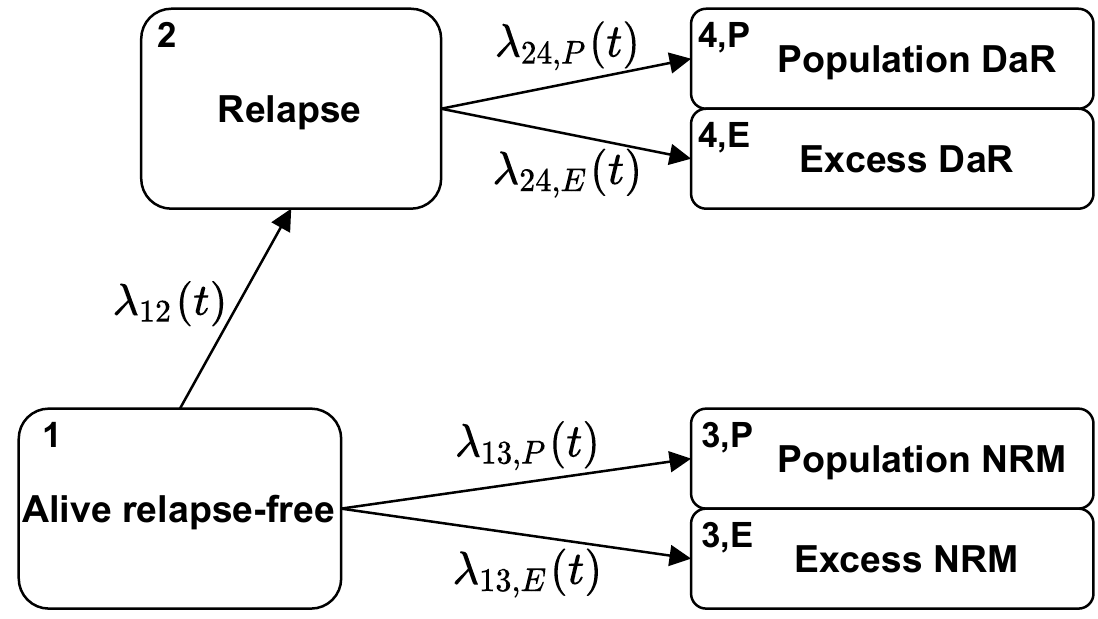}
\caption{Transition diagram for the extended multi-state model.}
\label{transmat2}
\end{center}
\end{figure}

We will now extend this basic multi-state model presented in Section \ref{mstatesec} with the relative survival approach introduced in Section \ref{relsurvsec}. Thus, we will split death states in a multi-state model into excess and population counterparts. To better illustrate the idea, we will extend the example given in Figure \ref{transmat}. By matching the observed data with population mortality data, we would like to estimate the amount of population and excess deaths, both before and after the intermediate event. Such a model is illustrated in Figure \ref{transmat2} -- although the exact cause of death is not present in the dataset (the observed data are the same as before), we show two additonal arrows and two additional death states by which we differentiate between excess and population death. 

We now discuss the quantities of interest and their estimation under this model. We start with the hazards. For transitions that do not reach death (for example the transition from Alive relapse-free to Relapse, see Figure \ref{transmat2}), the hazard is the same as in the standard multi-state model (Figure \ref{transmat}). It can again be estimated by the Nelson-Aalen estimator \eqref{NA}. On the other hand, each transition reaching death is now split into two (compare transitions 2 and 3 in Figure \ref{transmat} to the split transitions in Figure \ref{transmat2}). The hazards for these transitions have the same form as in Section \ref{relsurvsec}. The estimators follow from formulas \eqref{lambdaP}, \eqref{lambdaE} which we now write in general form. For a transition reaching death $h \rightarrow j$ the estimators of the cumulative hazards for the population and excess transitions, denoted by $(hj,P)$ and $(hj,E)$, are
\begin{equation}
\widehat{\Lambda}_{hj,P}(t)  = \int\limits_0^t \! \frac{\sum\limits_{i=1}^n Y_{h,i} (u) \mathrm{d} \Lambda_{Pi} (u) }{Y_h (u)}; 
\widehat{\Lambda}_{hj,E} (t) = \int\limits_0^t \! \frac{\mathrm{d} N _{hj} (u)}{ Y_h (u)} - \int\limits_0^t \! \frac{\sum\limits_{i=1}^n Y_{h,i} (u) \mathrm{d} \Lambda_{Pi} (u) }{Y_h (u)}.
\label{msmrelsurvfor}
\end{equation}

The estimators provided in formula \eqref{msmrelsurvfor} have the same form as in formulas \eqref{lambdaP}, \eqref{lambdaE}, the difference being that now through the at-risk process $Y_h(t)$ and counting process $N _{hj}(t)$ we can take into account additional competing risks (e.g.~relapse) and left-truncation. Transitions from the intermediate state to death are usually left-truncated since individuals experience the transition to the intermediate event at different timepoints after time 0. Partly as a consequence of the later entry in the state and partly because some demographic features might be associated with an increased risk for the intermediate event, demographic covariate values of those who experienced the intermediate event tend to be different from the whole group at baseline. These aspects are properly taken into account by continuously updating the population hazard based on the characteristics of the individuals observed to be at risk for each transition at each timepoint.	

Compared to the Nelson-Aalen estimator $\widehat{\Lambda}_{hj} (t) = \int_0^t \! \frac{\mathrm{d} N_{hj} (u)}{ Y_{h} (u)}$ which is a step function (jumps occur only at event times), $\widehat{\Lambda}_{hj,P}(t)$ as defined in \eqref{msmrelsurvfor} is continuous and piecewise linear, since $\lambda_{Pi} (u)$ is taken from mortality tables. Consequently, $\widehat{\Lambda}_{hj,E}(t)$ is defined as a difference of a step and a continuous function (and is càdlàg, like $\widehat{\Lambda}_{hj} (t)$). In practice, the estimation is done on small intervals (usually on a daily basis). However, $\widehat{\Lambda}_{hj,P}(t)$ and $\widehat{\Lambda}_{hj,E}(t)$ are then evaluated at event times only (like the Nelson-Aalen estimator $\widehat{\Lambda}_{hj} (t)$) and they are taken to be constant between event times. This procedure causes some bias, which is mostly negligible. The same approach is taken in \cite{Package} where it is also further discussed (see Section 3.4). Nonetheless, our \texttt{R} implementation allows for additional evaluation of the estimators at any choice of timepoints (e.g.~on a daily basis). An illustration of the estimation process is given in the Supplementary material, Section S2.

Having defined the estimates for transition hazards, we now turn to transition probabilities. Estimator $\mathbf{\hat{P}}(s, t)$ is again defined in matrix form \eqref{transprob}, whereby $\boldsymbol{\hat{\lambda}} (t)$ is extended using excess and population hazards for all split transitions. The product integral \eqref{transprob} then provides estimates, where for every split transition $h \rightarrow j$ we obtain estimates $\widehat{P}_{hj, P} (s, t)$ and $\widehat{P}_{hj, E} (s, t)$, whereas for all remaining transitions the definition of the estimator of the transition probabilities stays the same. In practice, transition probabilities are also estimated on small intervals but evaluated at event times only - a corresponding example is provided in the Supplementary material.

There are also other measures that are common in relative survival. One such measure is net survival \cite{mppnet}. However, it is questionable whether this measure is of interest for a multi-state model as its interpretation becomes less straightforward. 

\subsubsection{Variance estimation} \label{var.est}

We consider two approaches for estimating the variances of hazard and probability estimators - one based on the Greenwood estimator and the second one based on non-parametric bootstrap (simple resampling with replacement of individuals). 

We start with transition hazards: for non-split transitions we use the usual Greenwood estimator \eqref{var_greenwood} or bootstrap. To estimate the variances of the split hazards $\widehat{\Lambda}_{hj,E} (t) $ and $\widehat{\Lambda}_{hj,P} (t) $ we also consider two options. In the first, we assume that $\text{Var} (\widehat{\Lambda}_{hj} (t)) = \text{Var} (\widehat{\Lambda}_{hj,E} (t))$, thus we take $\widehat{\text{Var}}_{G} (\widehat{\Lambda}_{hj} (t))$ as the estimator of $\text{Var} (\widehat{\Lambda}_{hj,E} (t))$. In this case, one assumes that the population hazard $\widehat{\Lambda}_{hj,P} (t) $ is fixed and has variance zero. This assumption is often made because of the use of mortality tables which are deterministic. However, as is clear from formula \eqref{msmrelsurvfor}, the $\widehat{\Lambda}_{hj,P} (t) \text{'s}$ are not deterministic since $Y_h (t)$ brings some variability through the composition of the study population in terms of the demographic covariates (age, sex, year etc.). As a second option for evaluating variability we use the bootstrap, which will  help investigate if this additional assumption is reasonable.

Analogously, the variance of estimates for the transition probabilities is estimated using the Greenwood and bootstrap options.

\subsubsection{Confidence interval methods} \label{ci.methods}

We consider four confidence interval (CI) methods for transition hazards and probabilities. Although the model of interest is still the model introduced in the previous subsection, the methods are general and can be applied in any non-parametric multi-state Markov model. We denote by $\theta$ the transition hazard or probability at a given time for a given transition and $\hat{\theta}$ its corresponding estimate. By denoting $\widehat{\text{Var}}_{G} (\hat{\theta})$ and $\widehat{\text{Var}}_{boot} (\hat{\theta})$ to be the Greenwood and bootstrap estimators of the variance, respectively, we define the four CI methods.


 \begin{description}
   \item[(CI1)] \textit{Plain scale using Greenwood (plain.G):} a symmetrical $(1-\alpha)\%$ CI on the plain scale using the Greenwood estimator for the variance:
\begin{equation}
\hat{\theta } \pm \Phi (1-\frac{\alpha}{2}) \cdot  \sqrt{\widehat{\text{Var}}_{G}(\hat{\theta})},
\label{ci.plain}
\end{equation}
where $\Phi$ denotes the cdf of the standard normal distribution.
   \item[(CI2)] \textit{Plain scale using bootstrap (plain.boot):} a symmetrical $(1-\alpha)\%$ CI on the plain scale using the bootstrap estimator for the variance:
\begin{equation}
\hat{\theta } \pm \Phi (1-\frac{\alpha}{2}) \cdot  \sqrt{\widehat{\text{Var}}_{boot}(\hat{\theta})}.
\label{ci.plain.boot}
\end{equation}
   \item[(CI3)] \textit{Log scale using bootstrap (log.boot):} a $(1-\alpha)\%$ CI on the log scale (based on the delta method) using the bootstrap estimator for the variance:
\begin{equation}
\hat{\theta} \cdot \text{exp} \Big\{  \pm \Phi (1-\frac{\alpha}{2}) \cdot \frac{ \sqrt{\widehat{\text{Var}}_{boot}(\hat{\theta})} }{ \hat{\theta} } \Big\}. 
\label{ci.log.boot}
\end{equation}
   \item[(CI4)] \textit{Quantiles using bootstrap (q.boot):} a $(1-\alpha)\%$ CI where the limits of the CI are taken to be the
$2.5\%$ and $97.5\%$ quantiles of $\hat{\theta}$ over the bootstrap samples.   
\end{description}

\section{Simulations} \label{sim}

We analyse the characteristics of the developed methodology presented in Section \ref{mstaterelsurvsec} through simulations. First, in Section \ref{sim.details} we introduce all details regarding how the simulations were performed and in Section \ref{sim.results} we present the results. All simulations were done in \texttt{R}, version 3.5.2 \cite{R} using the packages \texttt{mstate} \cite{mstate11} and \texttt{relsurv} \cite{relsurv2018}. The main simulation functions have been added in a separate .\texttt{R}-file (\texttt{simulation\_functions.R}) in the Supplementary material.

\subsection{Defining the simulation study} \label{sim.details}

In this section we present the aims, data-generating mechanisms, estimands, methods and performance measures (i.e.~ADEMP) used for simulation. These are reported as proposed in the tutorial by Morris et al.~\cite{morris}.

\subsubsection{Aims} 

Our main goal is to investigate how the newly proposed estimators for the extension of multi-state models with population mortality perform in a wide range of simulation scenarios. We will evaluate their bias, standard errors (SE) and coverage probabilities.

\subsubsection{Data-generating mechanisms} 

\textbf{Transition diagram:} we use the same transition diagram as in Figure \ref{transmat2}. At time $0$, all individuals are alive  event-free. All individuals are followed up to a maximum of $10$ years.

\noindent \textbf{Generating event times:} for every individual we generate a trajectory, where the variables and event times are obtained in the following way:

\begin{enumerate}
\item Demographic variables $D$ are simulated independently for every individual: age (uniform), sex (binary), year (uniform). In all simulations, sex is distributed Bernoulli($0.5$) and year of diagnosis is taken Uniform[1990-01-01, 2000-01-01]. Age is also simulated with the Uniform distribution on a prespecified interval that will be adjusted for every simulating scenario so that the amount of population deaths is controlled.
\item We first have three event times for the following competing risks: relapse, population NRM and excess NRM. In practice, we only observe NRM and relapse, but in the simulations we also generate the underlying population and excess times to death. If an individual reaches state Relapse, times to population DaR and excess DaR are then simulated conditional on the time of relapse. All  events times are simulated independently \cite{meiramachado14}.
\item Times to death from population causes are simulated using population mortality tables for Slovenia (which are available in \texttt{R} - package \texttt{relsurv}, object \texttt{slopop}). For simulating the transition from Relapse to population DaR the demographic covariate values at time of relapse are considered.

\item Remaining event times (time to relapse, excess NRM and excess DaR) are simulated using $\lambda_i (t) = \lambda_0 (t) \cdot \exp(\boldsymbol{\beta}^{\top} D_i)$. 
\begin{itemize}
\item In the simplest case, we take the hazard to be constant (i.e.~we use the exponential distribution). In this case, the baseline hazard is constant and there are no covariate effects ($ \lambda_0 (t) =  \lambda_0$).
\item In some simulating scenarios, we take Weibull distributed event times for these transitions. In this case, $\lambda_0 (t)$ is the hazard of a Weibull distribution with parameters $a$ (rate) and $b$ (shape); $\lambda_0 (t) = a b t^{b-1}$. In this case, no covariate effects are considered. We further note that event times for the transition from relapse to excess DaR are simulated using a left-truncated Weibull distribution \cite{wingo_1989} where one conditions on the time of relapse; by this the Markov assumption is fulfilled.
\item Furthermore, we take simulation scenarios with a covariate effect (where $\boldsymbol{\beta}$ is not equal to zero). The covariate effect is included with respect to age, where age is first centered around 0.
\end{itemize}

\item Although we simulate event times for both population and excess transitions, only the minimum of these is included in the dataset as observed value.

\item The generated times are then subjected to censoring. Censoring times are simulated from the exponential distribution. The rate parameter of the distribution is modified for every scenario such that approximately 20\% of all individuals are censored at the end of follow-up time ($10$ years).
\end{enumerate}

\noindent \textbf{Simulation scenarios:} as mentioned, all population-related event times are generated using the same mortality tables, but population hazards vary between scenarios by adjusting the age distribution. Based on how the remaining event times are distributed and the magnitude of their hazards we define the five following simulation scenarios:

\begin{enumerate}
\item $exp.small$: for all transitions that are not population-related, the exponential distribution is used and the amount of population deaths is small.
\item $exp.large$: for all transitions that are not population-related, the exponential distribution is used and the amount of population deaths is close to the amount of excess deaths.
\item $weibull$: for all transitions that are not population-related, the Weibull distribution is used and the amount of population deaths is small.
\item $cov.e\hspace{-0.1em}f\hspace{-0.2em}f\hspace{-0.1em}.pos$: we allow for a covariate effect for certain transitions, the baseline hazard is again constant. The covariate effects of age on the hazard of having excess NRM and relapse are positive. No covariate effect is taken for the hazard of having excess DaR. 
\item $cov.e\hspace{-0.1em}f\hspace{-0.2em}f\hspace{-0.1em}.neg$: we allow for a covariate effect for certain transitions, the baseline hazard is again constant. The covariate effect of age on the hazard of having excess NRM is positive, whereas on relapse it is negative. No covariate effect is taken for the hazard of having excess DaR.
\end{enumerate}

%

Details on exact parameters can be found in the Supplementary material. We additionally assess the impact of sample size on the estimators by varying sample size: $n=500, 1000$ and $2000$ subjects. 

\subsubsection{Estimands} 

Our main measures of interest are the transition hazards and probabilities of the multi-state model. All values are assessed at $1, 2, 5$ and $10$ years, together with the corresponding standard errors and confidence intervals. All these measures have been defined in Section \ref{mstaterelsurvsec}.

Exact transition hazards and probabilities (the true values) were calculated by integrating with respect to the covariate distribution $D$, which was not straightforward since times to population death were generated using mortality tables (see Section S1 in the Supplementary material). The numerical integration was done in \texttt{R}.

\subsubsection{Methods}


We use the non-parametric approach outlined in Section \ref{mstaterelsurvsec} to estimate all transition hazards and probabilities, and we compare the performance of the variance estimators and the four different confidence interval methods.

\subsubsection{Performance measures}

\textbf{Bias:} let $\theta$ be the transition hazard or probability estimated at a given time for a given transition, whereas $\hat{\theta}_i$ is the estimate of $\theta$ for the i$^{th}$ replication of a simulation scenario. We define
$$\overline{\hat{\theta}} = \frac{1}{n_{sim}} \sum_{i=1}^{n_{sim}} \hat{\theta}_i. $$
Then the definition of absolute and relative bias follows:
$$\text{absolute bias} = \theta - \overline{\hat{\theta}}, \hspace{60pt} \text{relative bias} = \frac{\theta - \overline{\hat{\theta}}}{\theta}.  $$
Relative bias is reported in the simulation results. This allows for an easy comparison of results across different timepoints.

%

\noindent \textbf{Standard errors:} we compare the two approaches for estimating variances (Greenwood and bootstrap) as defined in Section \ref{var.est}. For the bootstrap $100$ replications are taken. We also calculate empirical standard errors, i.e.~the standard deviation across all simulation replications:

$$\widehat{\textrm{SE}} = \sqrt{ \frac{1}{n_{\text{sim}} - 1} \sum_{i=1}^{n_{\text{sim}}} (\hat{\theta}_i - \overline{\hat{\theta}})^2  }. $$ 

%

\noindent \textbf{Coverage probabilities:} $95 \%$ confidence intervals (CI) will be calculated for transition hazards and probabilities using the four methods introduced in Section \ref{ci.methods}. The estimated coverage probability (CP) is then the proportion of times the resulting CI contains the true estimand value.


We also use the coverage probability for calculating the number of replications $n_{\text{sim}}$ needed in the simulations. As in Section 5.3 in Morris et al. \cite{morris}, we require that the SE for the coverage probability equals $0.5\%$. Knowing that the expected coverage probability equals $95\%$, it follows that $n_{\text{sim}} = 1900$. We note that if we would require the SE for the coverage probability to be $1\%$ we would obtain $n_{sim} = 475$. In our simulations we round this to $n_{\text{sim}} = 2000$.


\subsection{Results} \label{sim.results}

\subsubsection{Bias}

\begin{figure}[ht!]
\begin{center}
\includegraphics[height=77mm]{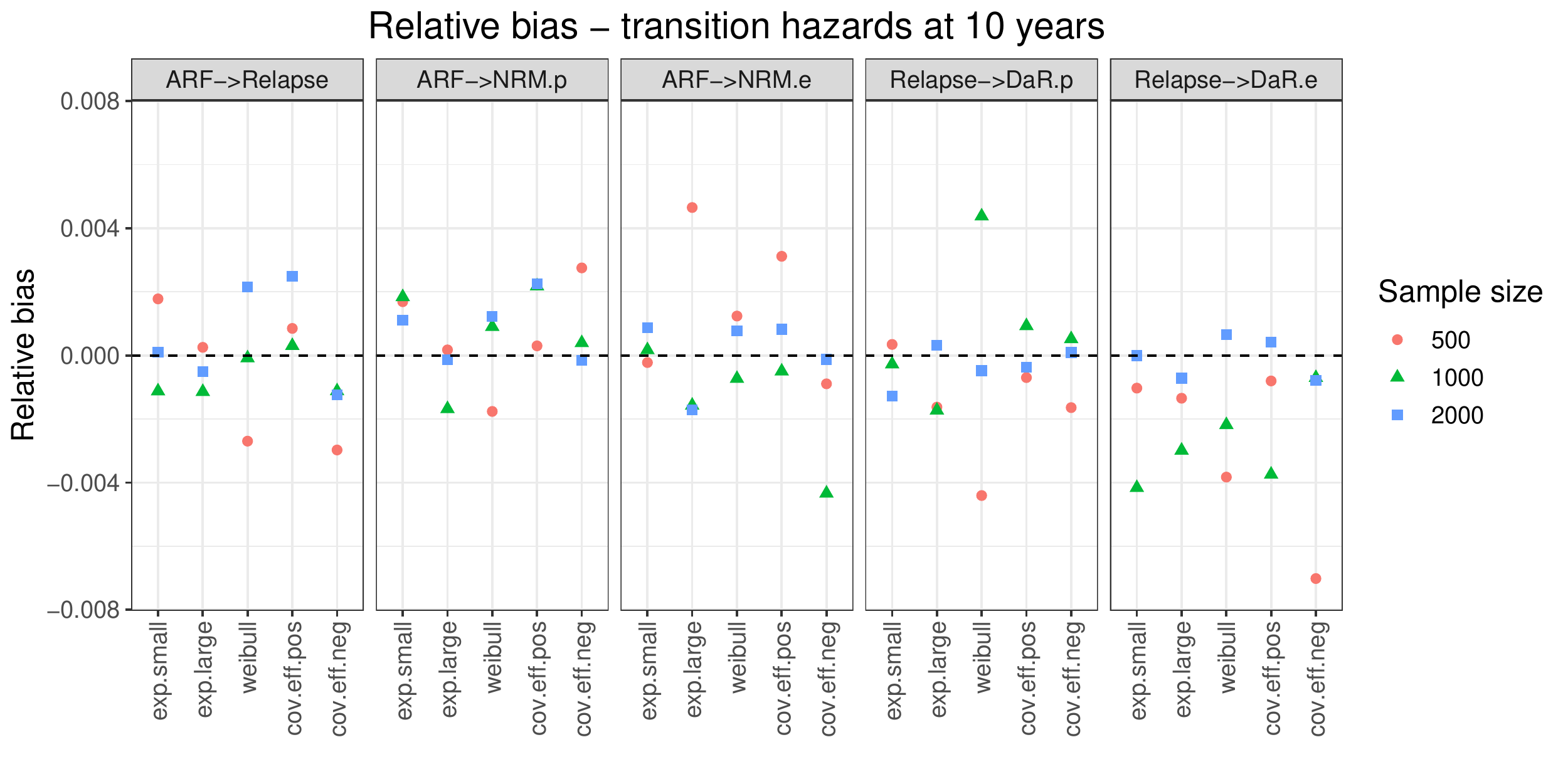}
\end{center}
\caption{Relative bias for cumulative hazards evaluated at end of follow-up ($10$ years). States: Alive relapse-free (ARF), Relapse, population NRM (NRM.p), excess NRM (NRM.e), population DaR (DaR.p), excess DaR (DaR.e).}
\label{relbias.haz}
\end{figure}

Bias was negligible in all scenarios. Figure \ref{relbias.haz} shows the relative bias for cumulative transition hazards for all scenarios evaluated at the end of follow-up ($10$ years). An equivalent figure for transition probabilities is given in Figure S4 in the Supplementary material. On average, relative bias was smaller when the sample size increased, which is expected. It seems that the relative bias was distributed around 0 and that behaviour was appropriate for all transitions and scenarios. For simulation scenarios with a covariate effect a slightly larger deviation from $0$ was seen, but when the sample size was large enough (e.g.~$2000$), the relative bias was close to $0$.

\subsubsection{Standard errors}

\begin{figure}[ht!]
\begin{center}
\includegraphics[height=77mm]{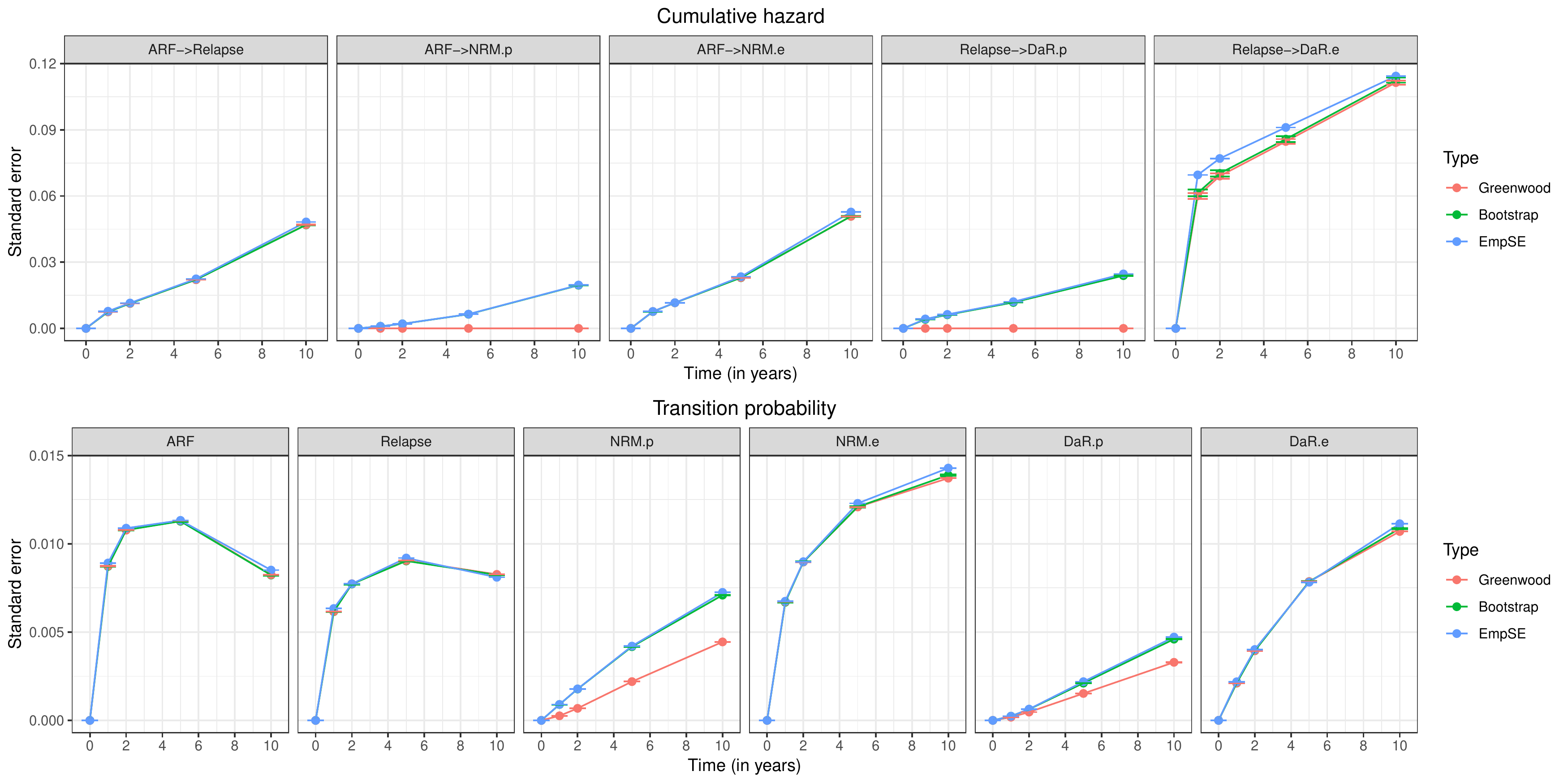}
\end{center}
\caption{Standard erorrs for cumulative hazards and transition probabilities for scenario $exp.large$ with sample size $2000$ (average over 2000 replications). States: Alive relapse-free (ARF), Relapse, population NRM (NRM.p), excess NRM (NRM.e), population DaR (DaR.p), excess DaR (DaR.e).}
\label{SEfig}
\end{figure}

 In Figure \ref{SEfig} we show standard errors for scenario $exp.large$ with sample size $2000$. Through this example we will explain  the crucial characteristics that were present in all of the simulation scenarios. The Greenwood option gave a zero estimate of the variance for population-related hazards and consequently, the estimated variance was also smaller for the corresponding transition probabilities. The bootstrap option better reflected the uncertainty in the estimates, as is confirmed by its values being close to the empirical SE. Furthermore, standard errors were smaller for population-related transitions than for their excess counterparts. When the amount of population deaths was increased, the SE of these population-related transitions increased as well.

When looking at the left-truncated transition going from relapse to excess DaR, we can see that both variance estimators for the cumulative hazards, Greenwood and bootstrap-based, gave evidently smaller SEs than the empirical SE. This occurred for all simulation scenarios. As further explained in Section S3 of the Supplementary material, this phenomenon does not only appear for the extended multi-state model but is already present in a basic multi-state model for left-truncated transitions. Solving this problem is out of the scope of this paper.

For transition probabilities (Figure \ref{SEfig}) both estimator options were close to the empirical SE, the population-related transitions being an exception. In this case, the bootstrap was again a reliable option. The standard errors for population-related transition probabilities increased when the amount of population deaths increased.

We find the bootstrap-based method a reliable option for estimating variances for all transitions in the wide range of scenarios we have explored. For some simulation scenarios (those with a covariate effect) there was a small difference between the bootstrap estimate and the empirical SE for samples of size $500$ and $1000$. However, these differences diminished for samples of size $2000$.

We additionally compared the standard errors for the split transitions to the standard errors for the observed transitions. In most cases, the empirical standard errors for the observed transitions were very similar to the ones for the excess-related transitions. This has been proven to hold asymptotically for transition hazards, i.e.~$\text{Var} (\widehat{\Lambda}_{hj} (t)) \approx \text{Var} (\widehat{\Lambda}_{hj,E} (t))$ \cite[p. 202]{andknjiga}. On the other hand, standard errors for population-related transitions were smaller than for their excess-counterparts and they were mostly dependent upon the amount of population deaths and variability in the distribution of the demographic variables.

\subsubsection{Confidence intervals}

\begin{figure}[ht!]
\begin{center}
\includegraphics[width=155mm]{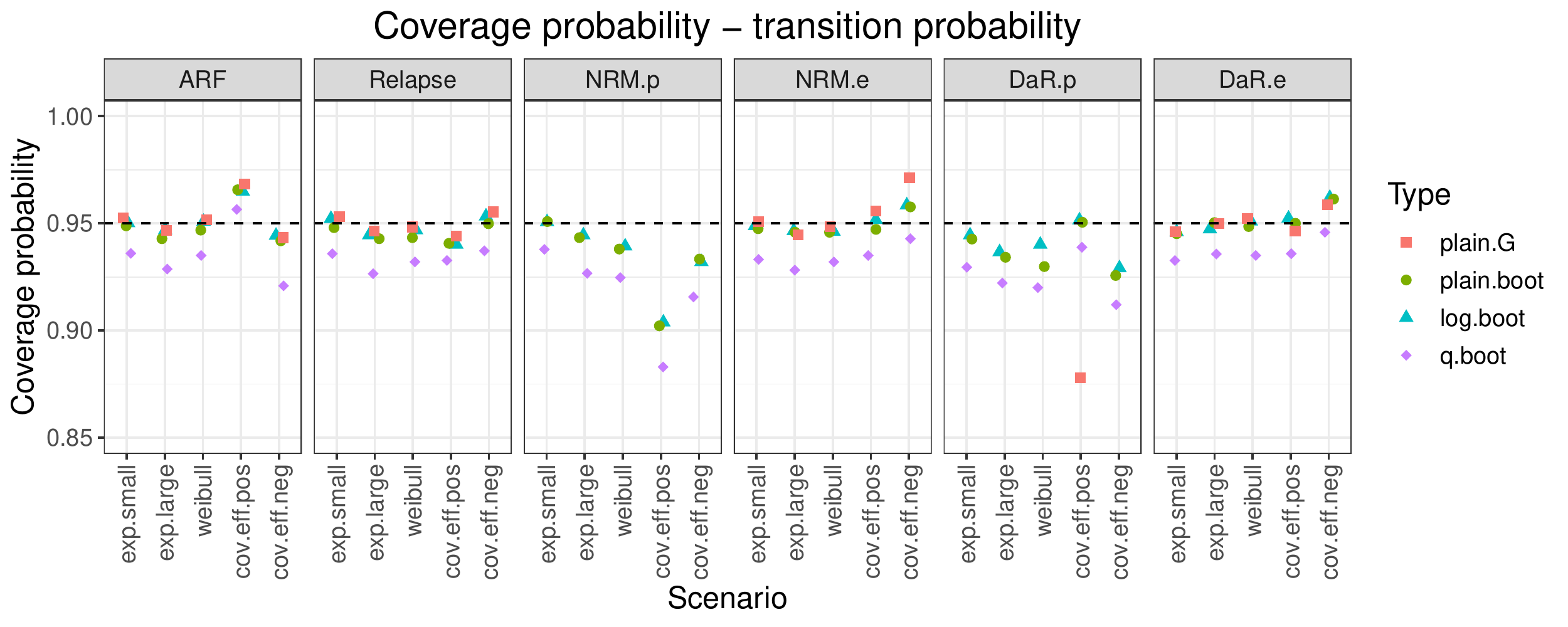}
\end{center}
\caption{Coverage probabilities for transition probabilities evaluated at the end of follow-up ($10$ years) (averaged across sample sizes). Method $plain.G$ gives inadequate coverage probabilities for population NRM and population DaR, many of which are smaller than $85\%$ and thus are not shown in the figure. States: Alive relapse-free (ARF), Relapse, population NRM (NRM.p), excess NRM (NRM.e), population DaR (DaR.p), excess DaR (DaR.e).}
\label{CI.prob}
\end{figure}

Figure \ref{CI.prob} shows coverage probabilities for transition probabilities at the end of follow-up. An equivalent graph for transition hazards is shown in the Supplementary material (Figure S5). On average, coverage probabilities were more stable and closer to the nominal value for transition probabilities than for hazards. By increasing the sample size the coverage probabilities also tended to increase and stay close to the nominal value (results not shown).

In almost all cases, $q.boot$ gave smaller CPs than the other methods. For population-related transitions, the $plain.G$ method gave inadequate confidence intervals - on the hazard level we got zero-width CIs, on the probability level they were somewhat larger but still far from the nominal value $95 \%$ (many of them were smaller than $85 \%$ and thus not shown in Figure \ref{CI.prob}). This is a consequence of the variance estimator that was taken to equal $0$ for these transitions.  

The remaining two methods $plain.boot$ and $log.boot$ which were based on bootstrap SE estimates behaved similarily in the simuations, but the latter method gave much better results for some cases. For left-truncated and population-related transitions the $log.boot$ method gave larger CPs which were closer to the nominal coverage - this was especially evident at earlier timepoints. An example of this is shown in Figure S6 in the Supplementary material. Thus, $log.boot$ is the preferred choice between the four introduced CI methods.

However, there were cases where even $log.boot$ gave lower coverage probabilities than expected; this was most evident for some left-truncated and population-related transitions for scenarios $cov.e\hspace{-0.1em}f\hspace{-0.2em}f\hspace{-0.1em}.pos$ and $cov.e\hspace{-0.1em}f\hspace{-0.2em}f\hspace{-0.1em}.neg$. This phenomenon occurred whenever the bootstrap SE estimate was lower than the empirical SE. 

\section{Illustration} \label{example}

The new methodology is briefly illustrated by a reanalysis of the data presented in \cite{leukemia2019}, which is taken as a motivating example in this article. The study population consisted of patients who had received a first allogeneic hematopoietic stem cell transplantation (alloHCT) for MDS or sAML between January 2000 and December 2012 and were recorded in the registry of the EBMT \cite{ebmt}. Further details on the selection criteria, patient characteristics and outcomes are available in \cite{leukemia2019}. 

Patients were followed since alloHCT and possible outcomes were relapse/progression (for the remainder of this section we call this relapse) or death where we again distinguish between non-relapse mortality (NRM) and death after relapse (DaR). Thus, a multi-state model with a transition diagram as in Figure \ref{transmat} is applicable, which can be further extended as in Figure \ref{transmat2}.
To better illustrate the usefulness of the methodology proposed, a subsample from the original dataset was taken, only including patients aged $\geq 60$ years at alloHCT and still alive relapse-free at 2 years after alloHCT (the landmark time). Population mortality is especially relevant for this group of older patients who have survived the first hazardous period after alloHCT. Data from a total of $753$ patients from $19$ European countries were included, with a median age of $63.9$ years, where $62\%$ of them were male and $38\%$ female. For this example, we restrict follow-up time to $8$ years.

\begin{figure}[ht!]
\begin{center}
\includegraphics[height=64mm]{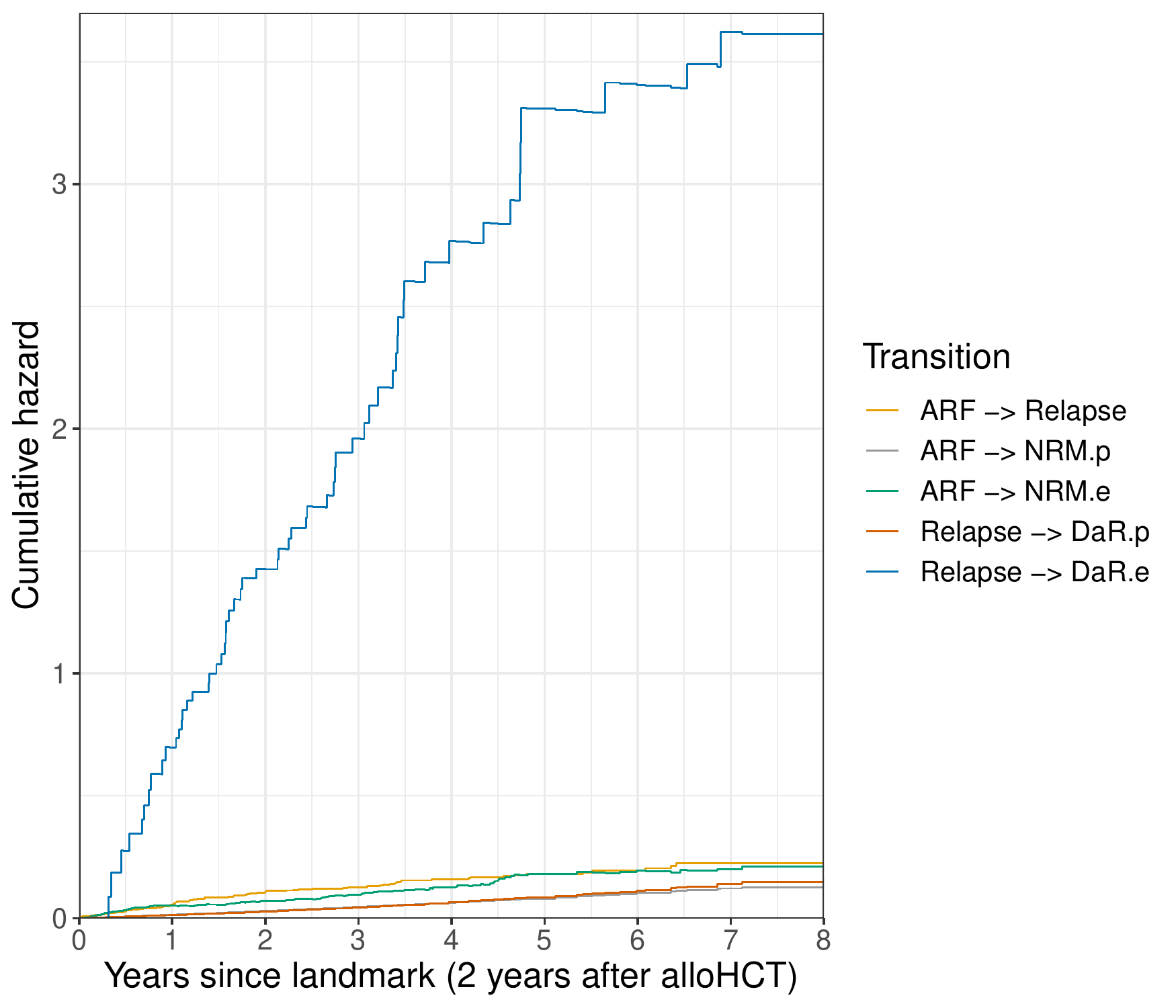} \hspace{-3mm}
\includegraphics[height=64mm]{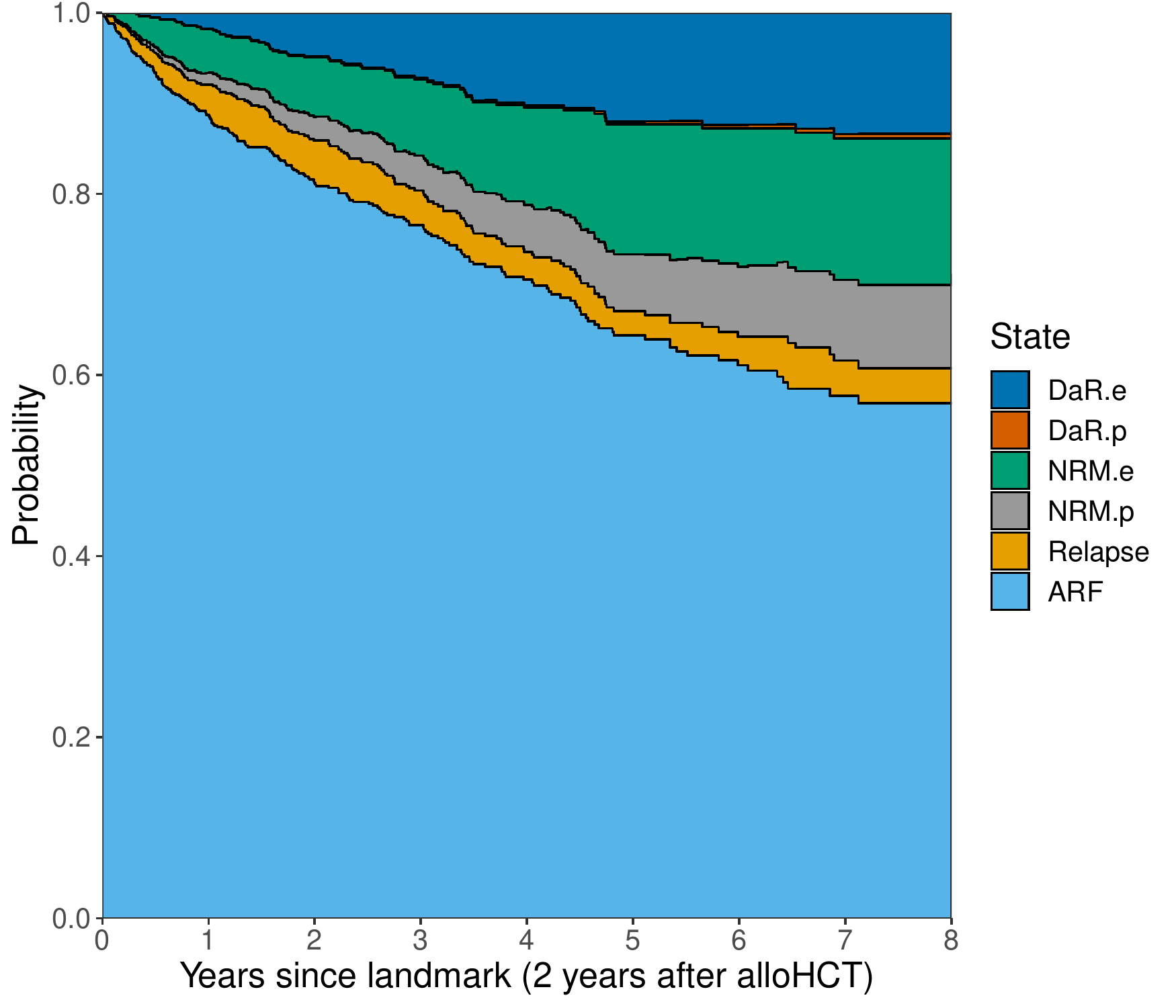}
\end{center}
\caption{Estimated cumulative hazards (left) and transition probabilities (right) for patients 65 years or older at alloHCT and alive relapse-free at the 2-year landmark time estimated in the EBMT dataset (as introduced in Section \ref{example}). Estimates provided up to $8$ years since the landmark time. States: Alive relapse-free (ARF), Relapse, population NRM (NRM.p), excess NRM (NRM.e), population DaR (DaR.p), excess DaR (DaR.e).
}
\label{examplefig}
\end{figure}

Figure \ref{examplefig} shows the estimated cumulative hazards (left graph) and transition probabilities (right graph) for the extended multi-state model, whereas in the Supplementary material, these estimated values are shown with the corresponding confidence intervals, see Figure S7 and S8. The huge cumulative hazard of excess DaR is a sign of the very poor prognosis after relapse. The cumulative hazard of experiencing relapse is for most of the time slightly larger than the cumulative hazard of excess NRM. The hazards for the population-related transitions are very similar, their difference only caused by the different composition of the risk sets in terms of demographic variables in the states Alive relapse-free and Relapse. 

The transition probabilities (right graph in Figure \ref{examplefig}) provide additional insight into the outcomes of the transplanted patients. The probability of being alive and relapse-free steadily decreases through time and reaches a probability of $56.9\%$ ($95\%$ $log.boot$ CI: $[52.1, 62.1]$) at $8$ years (all probabilities in this paragraph are reported at this time). The pattern shown by the hazards is propagated by the transition probabilities. The probability of dying due to population causes (population NRM or population DaR) equals $10.9\%$. The probability of population NRM equals $10.3\%$ ($95\%$ CI: $[9.7, 10.9]$), whereas its excess counterpart (the probability of excess NRM) is $15.1\%$ ($95\%$ CI: $[11.3, 20.2]$). On the other hand, the probability of population DaR is much smaller than of excess DaR: $0.6$ ($95\%$ CI: $[0.4, 1.0]$) and $13.3$ ($95\%$ CI: $[10.7, 16.5]$), respectively, which is due to the strong competing risk of excess DaR. The probability of being in the relapse state (and having no further event) is steady after two years reaching $3.9\%$ ($95\%$ CI: $[1.7, 8.8]$) at end of follow-up.

The clinical interpretation of these results is that death after relapse is almost identical to death due to relapse. In the past, only young patients were referred for alloHCT. For these younger patients (below the age of 60 years at the time of alloHCT) non-relapse mortality is almost exclusively treatment-related mortality (where treatment refers both to the alloHCT and to other treatments before and after). Yet age at HCT has continuously increased in the past decades, and the presumption that non-relapse mortality equals treatment-related mortality  is not valid for the older transplanted population. In our example, this is demonstrated by the fact that $40.6\%$ of NRM at 10 years after alloHCT for this landmark cohort can be attributed to population mortality. Thus, their treatment-related mortality can better be estimated by excess NRM.


The \texttt{R} package \texttt{mstate} \cite{mstate11} allows for the introduced extension. After the transition hazards are estimated using the function \texttt{msfit}, they are supplied to the newly developed \texttt{msfit.relsurv} function. This produces a new and extended \texttt{msfit} object containing the hazards for the excess and population transitions and their variances, calculated by the methods described in Section \ref{var.est}. The extended \texttt{msfit} object can then be supplied to function \texttt{probtrans}, through which transition probability estimates and their variance estimates are obtained. Furthermore, plain and log confidence intervals (as defined in Section \ref{ci.methods}) can be obtained using the \texttt{summary.msfit} and \texttt{summary.probtrans} functions. Thus the original package has been upgraded in such a way that only a small additional step is needed when coding (function \texttt{msfit.relsurv}). A full code example using \texttt{mstate} is provided in Section S4 of the Supplementary material.

%

%

%

\section{Discussion} \label{discussion}

In this paper, we have introduced an extension of a multi-state model containing death states where the exact cause of death is not known. We were able to account for death by other (population) causes in such a model using external population mortality tables. The relative survival theory has allowed for such an extension where we have solely focused on non-parametric estimation in a Markov setting. Thus, we have successfully integrated the basic relative survival idea (of splitting a death state in excess and population counterparts) in a general multi-state framework for any transition reaching death.

The present work focuses on transition hazards (through which the extension has been done) and transition probabilities, the latter being more easily interpretable for a non-statistical audience. The proposed estimators are defined in such a way that the information from mortality tables is properly taken into account. Based on our simulation study we can conclude that as expected, estimation bias is very minor.

By using the extended multi-state model we also have to assume the additive model  (formula \eqref{hazardAditivniModel}) for all split transitions. Thus, in order for  the excess hazard to be a proper hazard, it needs to be positive across the whole time interval. There are many practical settings where this does not hold (e.g.~\cite{antero-jacquemin2018}) because the population under study has a better life expectancy than  the general population, e.g., because primarily patients with a more robust general health might survive certain diagnoses. If the model is still applied even though it does not really hold, \textit{negative} cumulative hazards and transition probabilities are obtained, which could still give a useful message, if interpreted with caution. In such a case, negative cumulative hazards can only be obtained for excess-related transitions which then affect the transition probabilities of the excess and population death states.

Throughout the paper, we only split transitions to death states. However, the methods (together with the software implementation) allow transitions reaching intermediate events to be split as well. In this case, we would need adequate population tables for those specific transitions. For example, if we want to examine if patients with an autoimmune disease run more risk of a Covid-19 infection and death thereafter than the general population, a model could be developed with starting state Diagnosis, intermediate event Covid-19 infection and absorbing states Death before and after Covid-19 infection. Based on observed data for a cohort of these patients and population tables for the hazard of infection by Covid-19 and death after infection, the transition to the intermediate event could be split into a population and an excess part and similarly the transition from Infection to Death after infection could be split.  A different example is available in \cite{weibull2021} where a model for Hodgkin lymphoma patients with a disease of the circulatory system (DCS) as an intermediate event  is considered.

The suggested theory only uses a small part of the field of relative survival; it does not consider any other relative survival measures, e.g.~net survival or the relative survival ratio, and their extension to a general multi-state model. We leave this point for further research. We note that the crude probability of death (a common measure in relative survival) has its analogon in the split transition probabilities $\widehat{P}_{hj, P} (s, t)$ and $\widehat{P}_{hj, E} (s, t)$. 

For estimating the variance of the proposed measures we propose the bootstrap approach as best which was a suitable option in most of the simulation scenarios. The Greenwood estimator was also considered as a closed-form estimator. This required the additional assumption for the variance estimates of population-related transitions to be $0$. This assumption proved to be inadequate since the population-related transitions have substantial variability, whereas for the remaining transitions the Greenwood option was suitable in most cases. Instead of making such an assumption, a non-parametric variance estimator for population-related transitions would be of great use. However, such an estimator has not been developed (as far as we know) and thus this additional assumption was required. Developing such a variance estimator for the population-related transitions could find its use for many measures related to relative survival.

An additional problem with variance estimation of left-truncated transitions has been also pointed out (and further illustrated in the Supplementary material). This problem is inherent to a basic multi-state model already and is out of the scope of this paper but can be tackled in future work.

In the simulation study, we have also considered four confidence interval methods. The $log.boot$ performed best, and is therefore our recommended option. Even though there were cases when the coverage probability of $log.boot$ was evidently smaller than the nominal value, this method proved to be a reliable choice across simulations. Nevertheless, there is still room for improvement, either by defining a new confidence interval method or taking an even better variance estimator.

All work has been implemented in \texttt{R} and is free for use through the \texttt{mstate} package \cite{mstate11}. The implementation is based on functions from package \texttt{relsurv} \cite{relsurv2018} which has allowed for an easy and optimal usage of mortality tables which is otherwise not trivial. The current \texttt{R} implementation also allows for non-parametric estimation based on left-truncated observations in a standard survival setting (which has not been always considered in previous software implementations) and for future extensions to Cox-model based semi-parametric models. From what we know this is the first such implementation in \texttt{R}. Its use has been illustrated through an application. The Supplementary material shows self-contained code that can help the reader to apply this methodology.

To conclude, we believe that the proposed extension of a non-parametric multi-state model could find its use in a wide range of practical applications. Especially for interventions encompassing potential long-term beneficial and adverse effects offered to older patients with a remaining life-expectancy of several years or even a decade, taking into account population mortality as part of the different components of mortality is relevant. In this paper we have mostly focused on the theoretical aspects of this work. By conducting a thorough simulation study, we have further understood the properties of the estimators and come across additional challenges, some of which have to be dealt with in the future (e.g.~variance estimation for left-truncated transitions, a theoretical variance estimator for population-related transitions, an improved confidence interval method). We have devised an \texttt{R} application that relies on previous multi-state implementations in \texttt{R}. Thus by providing such a tool we believe that this work can be beneficial in many practical settings.

\subsection*{Acknowledgements}
Damjan Manevski is a young researcher funded by the Slovene Research Agency (ARRS). The authors acknowledge the financial support by ARRS (Methodology for data analysis in medical sciences, P3-0154 and project J3-1761). We thank the EBMT Data Office in Leiden for providing the MDS dataset.

\subsection*{Declaration of Competing Interest}

The authors declare that they have no known competing financial interests or personal relationships that could have appeared to influence the work reported in this paper.

\bibliography{reference}

\end{document}